# EDUCATING AND TRAINING ACCELERATOR SCIENTISTS AND TECHNOLOGISTS FOR TOMORROW

William Barletta*
*United States Particle Accelerator School and
Department of Physics, Massachusetts Institute of Technology
Cambridge MA 02139*
barletta@mit.edu

Swapan Chattopadhyay
*Cockcroft Institute,
Universities of Liverpool, Manchester and Lancaster and Daresbury Science & Innovation Campus
Warrington, Cheshire, WA4 4AD, United Kingdom*
Swapan@cockcroft.ac.uk

Andrei Seryi
*John Adams Institute,
University of Oxford, Royal Holloway University of London and Imperial College London
Oxford OX1 3RH, United Kingdom*
Andrei.Seryi@adams-institute.ac.uk

Accelerator science and technology is inherently an integrative discipline that combines aspects of physics, computational science, electrical and mechanical engineering. As few universities offer full academic programs, the education of accelerator physicists and engineers for the future has primarily relied on a combination of on-the-job training supplemented with intense courses at regional accelerator schools. This paper describes the approaches being used to satisfy the educational interests of a growing number of interested physicists and engineers.

*Keywords*: accelerators, education, USPAS, CERN School,

## 1. Introduction

Particle accelerators are essential instruments of discovery in fundamental physics, biology, and chemistry. Perhaps the best-known accelerator for fundamental research is now the 28-km Large Hadron Collider at CERN, Geneva that eclipsed the 6 km Tevatron as the world's highest energy particle collider. More numerous than machines for elementary particle and nuclear physics, and now more prolific scientifically, are the several tens of synchrotron light sources in nearly twenty countries around the world. These machines support a wide range of scientists studying condensed matter, atomic and molecular physics, chemistry and biology. At an even smaller scale and used for fundamental and applied research are ~180 machines for accelerator mass spectrometry.

Particle-beam-based instruments used in medicine, industry and national security form a multi-billion dollar per year industry. [1] For example, low energy electron beams are widely used for radiation cross linking and polymerization of materials to improve mechanical properties, such as tensile strength and scratch resistance, to increase the melting temperature of the material, or to increase its resistance to chemicals. Ion beam implantation of dopants is an essential process to the semiconductor industry. The economic value of goods treated with accelerator-generated beams is estimated to be more than 50 B$ per year. With respect to medical uses, accelerators account for roughly 100,000 radiotherapy treatments annually. Sales of accelerators account for more than 3.5 B$ per year of economic activity.

---
* Corresponding author and Director, United State Particle Accelerator School

In short, accelerators have become a hallmark of highly technological societies. More than 70,000 peer-reviewed papers with accelerator as a keyword are available on the Web. In spite of such a large economic and intellectual impact, only a tiny fraction of universities in the US offer any formal graduate education in accelerator science and its core technologies despite efforts by some national accelerator laboratories to expand that presence in major research universities.

Several reasons for this situation can be cited: 1) the science and technology of particle beams and other non-neutral plasmas cut across traditional academic disciplines. 2) Electrical engineering departments have evolved toward micro- and nano-technology and computing science. 3) Nuclear engineering departments have atrophied at many major universities although many now are rebuilding with a broadened emphasis on nuclear security and technology for next generation reactors. 4) With few exceptions, student interest at individual universities is not extensive enough to support a strong faculty line.  5) Funding agency support of university-based accelerator research infrastructure is insufficient to support the development of new faculty lines.

With some notable exceptions a similar observation also applies to universities worldwide. In 2009 a European consortium, Test Infrastructure and Accelerator Research Area (TIARA) conducted a web-based survey to quantify the level and modalities of accelerator training and education in Europe. TIARA reported [2] that although seventy-five organizations (universities and scientific laboratories) provide some level of training in accelerator science and technology, "there are only a handful of dedicated full-time formal training programmes in accelerator science." Still the number of persons in Europe receiving some training in accelerator science is large (nearly 1400), significantly larger than in the US.

As is true of many other modern technologies, the present generation of existing accelerators is continually pushing the limits of performance with respect beam characteristics (average current, particle energy, beam brightness), beam stability and reproducibility, machine reliability and time between servicing. Such requirements translate into increased size, complexity, cost and time to design and build[†] (or upgrade) large research accelerators. Consequently both commercial customers and research agencies are displaying greatly diminished tolerance for machine malfunction or failure to achieve full performance goals. Accelerators of the future will be even more demanding with respect to performance and cost-effectiveness.  Thus modern economic and technological environments are placing increased demands on the inventiveness and technical competence of the scientists and engineers who design, build and operate accelerators. In turn, those demands are increasing the pressure on interested stakeholders to develop and support stronger and more widespread educational and training programs.

This paper describes the primary formal and some informal approaches being used worldwide to train physicists and engineers in the science and technologies of modern accelerators. Although the material herein is far comprehensive on a country-to-country basis, it is representative of formal academic programs.

One can break down historical and present practice into a few categories:
- Self-instruction as part of one's experimental activities in an accelerator-based science,
- Apprenticeship training after formal education in physics or an engineering discipline,

---

[†] For example, the Large Hadron Collider at CERN took more than fourteen years to complete.

- Formal academic training in accelerator science and technology in a university program.
- Study at regional or international accelerator schools.

Frequently, accelerator professionals receive their education and training through a combination of these approaches. For example, those enrolled in formal university programs are frequently attendees at regional accelerator schools (as described in Section 3) taking one or more courses. At the post-undergraduate level[‡] (M.S., Ph.D. or equivalent) all of these approaches presuppose an educational foundation in physics or engineering.

## 2. University programs

For undergraduate and graduate students, a well-structured university program including hands-on experimental experience would seem to be the most direct entry to a career in accelerator science. The characteristic that best determines the health of a university accelerator program is the presence of viable faculty lines with a minimum of two tenure-track faculty who specialize in accelerator science and who can staff regular core course offerings. The field becomes slightly broader if one includes those physics and engineering faculties that have individual members with specialized interests in the field such as free electron lasers or plasma-based accelerators. In addition, some departments have nuclear and particle physics faculty who successfully place their students in national laboratories to do thesis research in accelerator physics and technology.

### 2.1. *Description of core university courses*

The most common foundational courses that are directly relevant to accelerator physics for both physicists and engineers are electromagnetism, classical mechanics, and applied mathematics (complex analysis, differential equations, Fourier transforms, special functions) at least the senior undergraduate level. For a Ph.D. level professional these topics should be at the graduate level and include a solid familiarity with special relativity and Hamiltonian formulation of mechanics. Electrical engineers are well served by having a course in waveguides, transmission lines and antennas – although many electrical engineering departments in the U.S. no longer offer such a course. Mechanical engineers should have a thorough familiarity with structural analysis, thermodynamics, heat and mass transfer. For both physicists and engineers, facility with statistics, numerical analysis and computational techniques is highly desirable.

For those universities that do offer graduate or undergraduate courses, the most common are an undergraduate "Introduction to Accelerators" and a graduate "Accelerator Physics." Undergraduate courses at the senior or upper-division level typically introduce the history and variety of types of particle accelerators, the physics of particle beams in linear and circular accelerators, transverse beam dynamics, acceleration of charged particles, longitudinal beam dynamics, synchrotron radiation, free electron lasers, collective instabilities and nonlinear effects. Sometimes selected laboratory measurements augment the lecture material. Such an undergraduate course is valuable in attracting students to future study and a career in accelerator science and technology.

The graduate course in accelerator physics would generally cover an introduction to the physics, technology, design, and operation of particle accelerators. Topics may include magnets for accelerators, the Hamiltonian formulation of single particle transverse and longitudinal motion, beam emittance, effects of linear magnet errors, chromatic effects and their correction, effects of nonlinearities, basic beam manipulations, RF systems, diagnostic systems, and an introduction to accelerator lattice design. Other topics are synchrotron radiation excitation and damping, beam-beam interactions, collective effects and instabilities. Computer-based calculations are frequently included in the syllabus.

---

[‡] The paper adopts the standard U.S. nomenclature. Graduate education (called post-graduate in the UK) refers to education towards an M.S. of PhD degree.

An essential aspect of a graduate-level educational program involves research carried out under the supervision of university faculty and senior research staff. In universities in which accelerator studies are multi-departmental, graduate students are usually admitted through a home department but may be required to pass a special examination overseen by a multi-departmental committee. In addition the student is typically required to write a thesis that is reviewed by an interdepartmental thesis committee.

The major research universities that are exceptionally well positioned to develop and educate the next generation of scientists and engineers are those which maintain a broad network of established educational and research partnerships with other universities and national laboratories, which have a world-class faculty, highly qualified students, and strong ties to local industry.

As relatively few universities have operating accelerator research facilities, a student's on-campus coursework can be greatly enriched through summer internships at collaborating institutions (such as national laboratories) and industrial affiliates. Internship programs provide students with hands-on training and mentoring from world-class accelerator scientists at active research centers. Participation in these distance-learning activities enhances the career development of students by 1) familiarizing them with evolving programs and career opportunities in accelerator-based science and technology, 2) establishing personal contacts with professionals in industry, the national laboratories and other universities, 3) emphasizing the integration of accelerator technology with the needs of accelerator-based physics, chemistry and biology, 4) strengthening career opportunities after obtaining their graduate degree. Moreover, long-term mentoring/monitoring relationships from experienced scientists in cutting edge research, often continuing during multiple years are recognized success factors in attracting and increasing the retention of students from underrepresented groups.

### 2.2. *Examples of national university-based programs*

Although university-based programs in accelerator science are relatively few in number, they have played a vital role not only in educating accelerator scientists but also in producing major advances in accelerator technology. It must be emphasized that many breakthroughs in accelerator science and technologies have been pioneered at the small number of universities with on-campus machines. Examples of inventions include the betatron, side-coupled linear accelerators, superconducting rf-accelerators, superconducting cyclotrons, pretzel beam orbits for high-luminosity collider operation, plasma-wakefield acceleration. Of course, such innovations require top-notch faculty lines[§] as well as highly talented students.

#### 2.2.1. *University programs in the U.S.*

The major research universities in the United States with structured programs that include graduate and undergraduate courses and that are producing PhD level physicists are the following (in alphabetical order): Cornell University, Indiana University, Michigan State University, Stanford University, University of California at Los Angeles, University of Maryland (College Park). Also initiating structured Ph.D. programs in accelerator science are Colorado State University, Massachusetts Institute of Technology, Old Dominion University (in affiliation with Jefferson Lab), and Stony Brook University (in affiliation with Brookhaven Lab).

To this list one may add a second group, namely, those universities with a single faculty member

---

[§] When a university commits to a tenured faculty line, it makes a sizeable, continuing commitment of funds (several million dollars in the U.S. That means that universities must expect faculty in the hard science and engineering to be able to secure sizeable research grants annually (~$300 per year in the U.S). Without sufficient opportunities from funding agency program offices, one cannot expect a sufficient cadre of world-class accelerator faculty in universities.

(either tenured or research faculty) whose primary research activity is accelerator science or multiple faculty with narrowly focused research activities: Duke University, Illinois Institute of Technology, Texas A&M, Northern Illinois University, the University of California at Berkeley, the University of Chicago, the University of Hawaii, the University of Southern California, the University of Texas at Austin and Vanderbilt University.

In the past some universities such as the University of Michigan and Columbia University had also produced accelerator Ph.D.'s, but they have none remaining in the pipeline as the single faculty advisor has left or is no longer accepting students. The lesson is that a single, interested faculty member has, at best, great difficulty in sustaining a university program. Figure 1[**] offers an historical look at those universities that have been the principal producers of Ph.D.-level accelerator scientists. Rather surprisingly the number of students (M.S. and Ph.D.) is a large fraction of (but in most cases is consistent with) the total historical production.

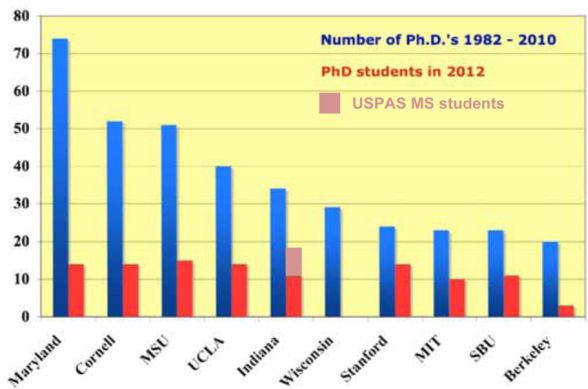

Figure 1: PhD's in accelerator science (blue) and present graduate students (red)

Even those universities with structured programs offer only two or three regular course listings in accelerator physics and engineering. The listed courses are generally an undergraduate and a graduate course in accelerator physics plus a regularly offered seminar style course in special topics. Cornell and MSU are the exceptions listing seven and five courses respectively. One MSU course at grants its students MSU-credit for any course taken at the U.S. Particle Accelerator School (USPAS). As the number of interested students may not exceed the respective university's minimum enrollment requirements, listed courses may not be offered every year. Consequently, all U.S. universities cited rely heavily on the USPAS to provide specialized academic coursework for their students. For this reason and as illustrated in Figure 2, the USPAS rubric of rigorous, for-credit courses hosted by major research universities is an essential aspect of formal accelerator education in America.

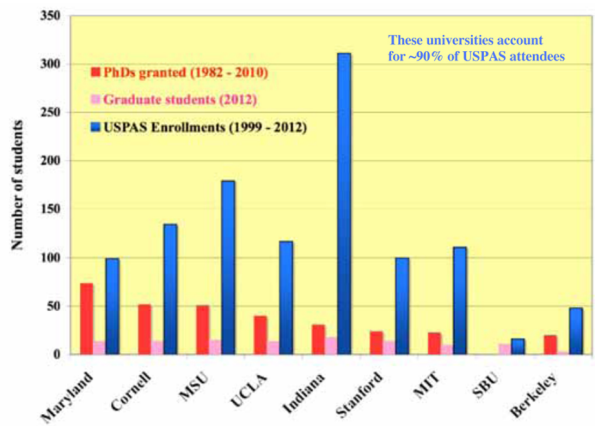

Figure 2: Student attendance at USPAS from the primary U.S. universities that produce PhDs in accelerator science

In addition to its highly successful Ph.D. program in accelerator physics, Indiana University in collaboration with the U.S. Particle Accelerator School offers the opportunity to earn a Master of Science Degree in Beam Physics and Technology. Students earn credit toward the Indiana University diploma at USPAS university-sponsored courses by selecting their USPAS course for Indiana University credit instead of credit from the host university. For each academic session, USPAS instructors are given appointments as visiting professors and the USPAS courses are added to the Indiana University curriculum. Award of a Master of Science Degree requires 30 hours of credit with a grade point average of B or above; a maximum of 8 credit hours may be transferred; some credits earned at previous USPAS courses may be eligible for transfer. There is a strict five-year limit to obtain the Master of Science degree. Generally, students may complete the Master's degree program within 3 years. At this time, international students are not eligible to enroll in the IU/USPAS Master's Degree Program. To date, Indiana has awarded IU/USPAS Master's Degrees to seven students. Presently there are seven active students in the Master's program.

---

[**] Both sets of data were provided to the author by the universities cited. Note that the expected number of PhDs produced annually is roughly 20% of the present level of students.

A crucial part of any student's training is the opportunity to participate in cutting edge accelerator research programs. Given top-quality faculty supervision, students can do accelerator research in areas that are central to an institution's accelerator development program. An outstanding example of such work is the optimization of superconducting rf-cavity structures as part of Cornell's ERL research program. At MSU a large number of students play a strong, active role at the National Superconducting Cyclotron Laboratory (NSCL). NSCL graduate research topics include superconducting rf-cavity design, modeling, and measurement techniques, SRF-related material science, cryogenics research, high-intensity ion-source development, large-dynamic-range beam instrumentation – in short, a broad reach of opportunity for both physicists and engineers. At UCLA, the extensive, world-class experimental program in plasma accelerators (in both the Physics and Electrical Engineering Departments) and in free electron lasers has produce a new generation of intellectual leaders in advanced acceleration techniques, computational accelerator physics, and X-ray source design for photon science. Unlike MSU, UCLA does not have a major accelerator[††] on campus; therefore, university faculty members have forged a strong, continuing partnership with the SLAC National Accelerator Laboratory.

At the University of Maryland, the novel electron-model storage ring (UMER) has played a vital role in advancing the understanding of the transport of space charge-dominated beams and has produced a substantial fraction of the PhDs in accelerator physics and engineering from U.S. universities. U.S. government investment in a few more small-scale research accelerators at universities would pay large dividends to America's large, accelerator-based science programs.

*Laboratory-affiliated programs in the U.S. and Canada*

The Department of Energy national laboratories with accelerator user-facilities have a powerful interest in assuring an adequate supply of highly trained accelerator scientists and technologists. They must and do play an essential active role in the education and training of accelerator scientists and engineers. Summer Undergraduate Laboratory Internships Student (SULI programs) are one way; providing instructors and financial support for USPAS sessions is another; providing research opportunities to graduate students for thesis projects is a third. In addition, several national laboratories have entered into educational agreements with one or more major research universities.

To encourage and expand Ph.D.-level education in accelerator science, in 1985 Fermilab established its Joint University-Fermilab Doctoral Program[‡‡] in agreement with major research universities. This program offers thesis research topics using Fermilab research facilities and expertise available at Fermilab. Fermilab funds student salaries, and provides day-to-day supervision of the student while the university thesis supervisor oversees the student's academic progress toward the Ph.D. The program currently has 10 students enrolled and has trained 41 Ph.D. students since its inception.

Fermilab is also in the process of building the Illinois Accelerator Research Center (IARC) on the Fermilab campus. IRAC will include educational space for collaboration with local universities to train scientists, engineers and technical staff in accelerator technology.

---

[††] UCLA does have a small machine, PEGASUS, which is very useful in the UCLA training program.

[‡‡] http://apc.fnal.gov/programs2/joint_university.shtml

More recently the Brookhaven National Laboratory (BNL) has established CASE[§§] – the Center for Accelerator Science and Education (CASE) in collaboration with Stony Brook University. CASE aims at training graduate and post-doctoral students as accelerator scientists through a combination of academic courses and hands-on experience with accelerators. To attract students to graduate programs, CASE offers undergraduates an introductory accelerator course, accelerator laboratory work and summer research opportunities at Brookhaven. CASE is composed of faculty from the BNL Collider-Accelerator Department and Accelerator Test Facility and the Stony Brook University Department of Physics and the School of Engineering.

In 2008 Old Dominion University (ODU) and the Thomas Jefferson National Accelerator Facility (JLab) joined forces to establish the Center for Accelerator Science in ODU's Department of Physics. Two years earlier ODU had launched an instructional program in accelerator physics with the help of JLab. The Center is now in the process of establishing the formal requirements for a Ph.D. degree in accelerator physics. In addition to courses offered on the ODU campus, students will be able to take all USPAS courses and receive direct ODU credit.

Expanding opportunities for PhD-level education will not bear fruit if talented undergraduates in physics and engineering are uniformed of and not attracted to them. As a first step to attract high quality students, the Argonne National Laboratory, Fermilab and the USPAS instituted the Lee Teng Undergraduate Internships[***] in FY 2008. Each year between ten and twelve students are selected for the program. Lee Teng Interns[†††] should have just completed their junior year (or for exceptionally talented students, their sophomore year) prior to the summer of the internship. All interns receive a scholarship to take the USPAS course, "Fundamentals of Accelerator Physics." They then complete an eight-week research project at ANL or FNAL under the supervision of a mentor. The project should have sufficient complexity that it can serve as the basis for an undergraduate thesis if that is required by the student's home university. The mentors remain available to guide the student through graduate school application and/or to advise the student in preparing the senior thesis. Moreover, participating in a summer research internship substantially raises the chances of a student's being selected to receive a government-sponsored fellowship.

In Canada the University of British Columbia, in collaboration with the TRIUMF laboratory and the University of Victoria, has initiated a multi-pronged approach in which graduate students conduct their Ph.D. thesis research at TRIUMF under the mentorship of TRIUMF scientists. Presently two accelerator courses are offered in alternate years: 1) Physics and Engineering of Particle Accelerators: Electrons, and 2) Physics and Engineering of Particle Accelerators: Protons and Ions. Here again students must rely on on-the-job training and specialized at regional accelerator schools.

### 2.2.2. *Accelerator training in Europe*

The TIARA report [7] offers an extensive analysis integrated on a country-by-country basis. The data in this section is based solely on that report. In addition TIARA identifies 11 universities that offer 100 contact hours or more of instruction in accelerator science. These institutions are the following:

---

[§§] http://www-case.physics.sunysb.edu/wiki/index.php/Main_Page
[***] http://www.illinoisacceleratorinstitute.org/

[†††] Ten Lee Teng Interns are selected each year. The selection committee not only chooses the awardees but also matches them with the mentors at each laboratory. The author has been pleased to teach the Lee Teng interns each summer at the USPAS session.

1. University of Manchester
2. Universitat Autónoma de Barcelona
3. IKP, TU Darmstadt
4. Institut für Kernphysik der Johannes Gutenberg-Universität Mainz
5. University Paris-Sud
6. IKP, FZ Jülich
7. DELTA, TU Dortmund
8. INFN-Milano & Università degli Studi di Milano
9. EPFL: Swiss Institute of Technology Lausanne
10. Università di Roma "La Sapienza"
11. Hamburg University

In 2011 these eleven universities accounted for 181 students who received 100 contact hours of training. Of nearly 1200 students who received at least 10 contact hours of formal training in accelerators, ~35% were undergraduates. Integrated over all institutions, 198 PhD students receive at least some training in accelerator science. Of all institutions which offer accelerator training of Ph.D. students 45% offer formal examinations on accelerator science course work.

2.2.3. *University programs in the U.K.*

In the England and Scotland, university consortia have acted together to overcome the limitations of resources and enrollment that hamper education in accelerator science and technology. The result is a set of exemplary university-based programs.

The Cockcroft Institute of Accelerator Science[‡‡‡] and Technology (CI) and the John Adams Institute for Accelerator Science[§§§] (JAI) are centers of excellence for innovative accelerator technology. They also provide research opportunities, and academic training in accelerator science and technology for students.

Established in 2004, the two institutes have become internationally recognized centers of accelerator science. Their role in educating the next generation of scientists has boosted the UK's impact in this area, helping to address a worldwide shortage of accelerator scientists and technologists. At both institutes the faculty, research staff and students collaborate on research programs at the forefront of accelerator science, spanning national and international facilities and projects. They also promote accelerator applications in science, medicine and industry.

Both institutes have well developed connections with industry and local communities through active outreach programs. These programs bring scientific ideas closer to practical applications and inspire younger generations toward careers in science and engineering.

The Scottish Universities Physics Alliance (SUPA) pools physics research and post-graduate education for doctoral students registered in eight Scottish universities: Aberdeen, Dundee, Edinburgh, Glasgow, Hariot Watt, St. Andrews, Strathclyde and West of Scotland. Approximately 120 students join the program each year.

*2.2.3.1 Cockcroft Institute Education and Training*

The primary goal of the CI education policy is to provide graduate students with a comprehensive education in both fundamental accelerator science and practical engineering. Cross training is encouraged; "hands-on" RF engineers benefit from understanding the elements of lattice design; theoreticians learn about of the challenges of an ultra-high-vacuum system.

Much of the training to date utilizes the extensive expertise available 'in-house' as well as invited lecturers from elsewhere in the UK, Europe and the U.S., providing valuable cross fertilisation between the CI and external organisations. Likewise, CI staff members deliver lecture courses at the John Adams Institute, the U.S. Particle Accelerator School and the CERN Accelerator School.

Over the last three years, 243 contact-hours of lectures have been offered, of which 145 were devoted to accelerator science or associated mathematics and 98 to accelerator engineering and technology. From early 2007, the CI lectures courses have been web-cast in real-time, and many organisations including the Diamond Light Source, RAL, e2v in Chelmsford, Essex, use the web-cast lectures to provide staff training.

---

[‡‡‡] The Cockcroft Institute website is http://www.cockcroft.ac.uk
[§§§] The JAI website is http://www.adams-institute.ac.uk/

This pattern has established a template for the education program for the future. Most of the topics originally envisaged for inclusion in the education program have been covered many times. Growth in the number of students requiring lectures will be driven by the ongoing enrolment of CI academic staff. New acceleration methods and technologies with novel problems and priorities have emerged in response to diverse demands from all areas of science. The CI education program continues to respond and rise to these challenges.

The CI lecture program has been established primarily, but not exclusively, for the Institute's graduate students. Depending on the topic, certain lectures or courses have attracted a wider interest amongst associated faculty members and Daresbury accelerator staff, as well as from other universities.

Currently there are 46 students enrolled in Ph.D. studies at the Cockcroft Institute. The institute has students based at the Cockcroft Institute, their respective University, at CERN and at the DESY laboratory.

*Course structure*

The core courses are repeated annually, but are not necessarily presented by the same lecturers. The aim is provide an introduction to accelerator physics, together with a detailed set of courses on essential topics, in the autumn term. The CI is able to rely on well-established experts in these areas, based at the Accelerator Science and Technology Centre (ASTeC) and at the three universities of the CI, to provide students with high quality lectures. In the spring term introductory lectures continue, together with a series of lectures aimed at an intermediate level of education – such as beam diagnostics for example. The summer term focuses on specialized topics and allows students to be exposed to international experts in the field.

As indicated, the lecture courses are divided into three categories:

• Fundamental Accelerator Physics, Engineering and Technology covers core topics that are central to understanding the design and operation of particle accelerators. They are regarded as essential material for all registered for post-graduate courses. CI intends to provide instruction in these subjects to all students during their first year of activity in the Institute.

• Intermediate and Advanced Accelerator Physics, Engineering and Technology covers topics which are central to the full understanding of accelerator behaviour but which, because of their nature, can best be assimilated after exposure to the basic courses and some experience in working on accelerator topics. They will be covered on an approximate three-year cycle. The majority of CI students are expected to attend these courses, so that upon graduation they will have a comprehensive understanding of major topics.

• Advanced and Specialist Topics may be individual lectures or courses focusing on niche areas which, depending on the research being carried out in the Institute, will be of great significance to some but less relevant to others.

The CI's education and training committee has considered the level of mathematics and physics experience and education that entering graduate students could reasonably be expected to possess. The basic course has been expanded to provide tutorials in a number of areas of maths and physics that are fundamental to accelerator science.

The development of complementary skills is regarded as a mandatory component of the training of graduate research students. Key requirements, which include the development of communication, networking and research management skills, have the aim of improving the participants' career skills. Sessions focus on project management, presentation skills and scientific writing. The course concept has been extremely successful and has become established as standard for all graduate research students in the School of Physical Sciences at the University of Liverpool.

*Coordinated programs*

The Cockcroft Institute coordinates the European Union funded Initial Training Networks DITANET, oPAC and LA$^3$NET. Training within these networks is composed of research, complemented by local training by the individual institutions, often in close collaboration with partners, and network-wide events, such as schools, topical workshops, or conferences. Projects join universities, research centres and industry to provide training to 60 early-career and experienced researchers. The aim is to give participants an excellent base for their future careers and to maximize their employability.

Over the past four years this activity has included organizing two international schools on beam instrumentation and diagnostics with around 70 international participants each, eight topical workshops in focused areas of beam diagnostics, such as longitudinal beam profile measurements or beam position monitoring with 30-50 participants each, an international conference on beam instrumentation and a major symposium in this research area. These events were open to the worldwide accelerator community with many participants coming from overseas.

*Outreach activities*

The CI's engagement with the public and schools involves organizing open days and schools master classes in the Institute, making visits to local schools, offering work experience, attending career fairs and hosting stands at national science fairs. Through its outreach program CI aims to

• Encourage students to study physics and engineering at university.

• Widen knowledge of the widespread use of accelerators in society beyond particle physics.

• Inform the local community of what is done at the Institute.

One example is the CI participation in the UK Particle Physics Master Class (PPMC) in which about 100 school children from the local area participate in interactive activities and lectures about particle physics. The Daresbury master class traditionally has had an accelerator physics slant As of 2012, plans are being made to run a separate accelerator physics master class to target students interested in particle accelerators. The activities will include beam dynamics computer simulations, the design of a synchrotron light source and measurement of the beam energy from the ALICE photo-injector from the bend caused by a dipole magnet.

In 2008 and 2010, CI organized and participated in a daylong symposium of SciTech events in London with participation by government and industrial sectors, showcasing accelerator science and technology. In 2010, CI also participated, together with JAI, in the annual Royal Society Summer Science Exhibition in London broadcast by BBC.

The JAI runs a similar outreach program, including master classes, the APPEAL program for school A-level teachers, the ACCELERATE program, participation in Science Festivals, etc.

*2.2.3.2. John Adams Institute programs*

The JAI educational mission at both the graduate and undergraduate levels involves students in cutting-edge research programs at state-of-the-art facilities. JAI's combined expertise at Oxford, Royal Holloway University London (RHUL) and Imperial College London (ICL) provide strong academic and technical support to students in areas as diverse as accelerator physics, beam diagnostics, lasers, plasma physics, laser-plasma interactions and computational techniques. The integrated training program in these areas addresses the UK's demand for a scientifically skilled workforce to sustain world-class research and to support the growth of a high-technology economy.

The JAI fully exploits the full range of accelerator-related expertise present across the three universities. This partnership allows students to follow a training program focused on their research whilst keeping abreast of the key concepts and wider developments in accelerator and plasma science and technology.

*The graduate program*

The graduate training program consists of a lecture course in Accelerator Physics taught by JAI faculty with support from guest lecturers, notably from CERN, the Cockcroft Institute and the Rutherford Appleton Laboratory (RAL). Students from both Oxford and RHUL enrolled in the PhD program attend these lectures. The current ICL graduate program in plasma physics, which covers both basic plasma physics and advanced topics, including plasma-based accelerators, will be integrated into the JAI program through the use of web broadcasting technology.

The core Accelerator Physics syllabus covers the basics of accelerator science and technology in the first two academic terms of the PhD program. In this way the students will have acquired a solid technical background before they start their research work, typically during the third term. This course is complemented by a number of supporting courses that deal with Electromagnetism for Accelerators, Hamiltonian Dynamics and Applications of Accelerators. The students can also attend other courses provided within the PhD training at the three universities that cover Computing, Statistics, Particle Physics and other advanced topics.

An important feature of the Accelerator Physics course, developed over the last five years, consists of a design project that students must complete at the end of the second term of their first year. The team of students is assigned to design an accelerator and its main subsystems with minimal initial input. The project is usually related to actual projects under development at other institutions. As part of the design project evaluation, the students present their work in an open seminar at the JAI and also at CERN. The CERN visit includes a tour of some of the accelerator facilities. In future, the visit to CERN may be lengthened to include further training by working alongside expert accelerator operation teams during machine runs. The design project has proven to be very successful in that it exposes the students to real-life problems at an early stage in their training and challenges them to work on difficult topics collegially. Responses from both students and scientists working on the selected projects have been unanimously positive.

The graduate lecture course is complemented by a series of JAI seminars, which feature world-class experts in accelerator science presenting their latest results on a wide range of topics in accelerator science and technology. The talks are advertised on various accelerator networks and are broadcast via Webex. The talks are also filmed and both slides and films can be downloaded and viewed on the JAI website where a full list can be found. The lectures are accompanied by a half-day visit of the speaker where more in-depth discussions with JAI staff take place. These lectures and seminars help the students make the transition from academic studies to research.

Each of the three universities offers courses on transferable skills such as Research Management and Critical Thinking, Personal Effectiveness, Written and Oral Communication Skills, Team Working, Research Skills and Techniques, Career Development, Entrepreneurship, Ethics, Teaching and Advanced Information Technology. Every student may access all modules available across all three institutions.

*Undergraduate education*

JAI also maintains a concerted effort to ensure that highly qualified and motivated undergraduates are ready to commence graduate training and research in accelerator science. The undergraduate program in accelerator science is attended by an average of 15 students per year at RHUL and Oxford. These courses will be complemented by the successful, long-established undergraduate course at ICL on plasma physics that attracts more than 100 students. The Accelerator Physics course consists of a simplified introduction to accelerator science and technology, suitable for students in the second or third years of their undergraduate degree. The undergraduate course at RHUL also provides guest lecturers from accelerator laboratories at local facilities. The exposure of the students to real life problems in accelerator physics has been widely appreciated by the students and has been extended to the courses offered at Oxford.

The JAI also hosts a number of internships for undergraduate summer students in their second or third years. A many potential PhD candidates have opted for accelerator projects in recent years, thereby attracting new students to the field. The JAI also provides a summer program for students from foreign universities such as Paris VII, hosting a total of about ten self-funded students during the last two years.

*Financial support*

The main source for support of graduate students has been the Science and Technology Facilities Council (STFC) studentship program. Of more than 30 students presently enrolled in the JAI PhD program, about 50% are funded by STFC. In recent years several Ph.D.s have been funded jointly with CERN's Doctoral Student program. Another significant source of recent funding is a Ph.D. bursary scheme offered by the Thai government, to support their best students, selected by a national competition, to study accelerator physics. Other sources include CASE studentships (co-sponsored by industry) and various scholarships offered by the individual universities. Siemens Technology has recently funded one Ph.D. student fellowship.

The ICL Plasma group has funded students through Engineering and Physical Sciences Research Council *(*EPSRC) project grants and the EPSRC Doctoral Training Account. All of the recent graduates have continued in the field of laser-plasma accelerators or laser-plasma interactions at universities in the UK and U.S. Students recruited into the accelerator research activity of the HEP Group at ICL have been funded through a number of sources including STFC quota awards, project grants, CASE awards and the EPSRC industrial CASE studentship scheme.

*Enrollment statistics*

Since 2005 the JAI has enrolled an average of five students per year. Since 2008 eleven students have graduated, all of whom are now employed, mostly in accelerator physics (7 students), two in finance and two in industry. Five students are currently writing up and are expected to graduate this year. The Ph.D. topics reflect the structure of the core activities of the Institute and were initially focused on the high-energy physics program (ILC, NF, MICE, etc.) and later extended to include light sources (Diamond and 4th generation light sources). In 2011 there was a significant increase in the student intake due to the inclusion of laser-plasma wakefield acceleration and the compact-light-source program into our core activities. The extension of the JAI to include ICL will strengthen the new direction towards laser plasma research.

*2.2.3.3. Scottish University programs*

Two courses in accelerator physics are part of the Nuclear and Plasma Physics theme of SUPA and are linked with the experimental research in laser-plasma acceleration and free electron lasers at the University of Strathclyde. The introductory course covers the history and applications of accelerators as well as core topics of beam characteristics and diagnostics, longitudinal and transverse dynamics, non-linear dynamics such as resonant phenomena, RF cavities and waveguides. The second, advanced course addresses the topical research in laser plasma interactions, laser-plasma acceleration and plasma based radiation sources.

SUPA also seeks to increase its commercial engagement with industry. This mission aims to generate revenue for SUPA and to facilitate the commercial exploitation of research to the benefit of the economy of Scotland and beyond.

### 2.2.4. *University program in Russia*

Two notable educational programs in Russia can be cited: the university program at Novosibirsk State University (NSU) and the program conducted by the Joint Institute for Nuclear Physics (JINR) at Dubna. The latter is conducted through the JINR University Center.

#### 2.2.4.1 The Novosibirsk university program

With its very close ties to the Budker Institute of Nuclear Physics (BINP), NSU has long had an extensive program of formal educational conducted by a formidable Department of Accelerator Physics[****], the graduates (and former faculty) of which work in almost all the largest world centers of high-energy physics and photon science. Although the department was formed as an independent unit in 1992, before then training in accelerator physics was conducted in the Department of Nuclear Physics, which was transformed in 1985 into the Department of Elementary Particle Physics. The deaprtment is complemented by the Department of Radiophysics.

The leaders of accelerator physics at NSU have included two BINP directors: Acad. G. I. Budker and Acad. A. N. Skrinsky. Not surprisingly the main areas of specialization of students in the department are closely aligned with experimental programs at BINP: conducting colliding beam experiments on VEPP-4, participating in the creation of the new electron-positron collider (VEPP-2000), creating a new electron-positron injection complex (VEPP-5), developing a free electron laser, and delivering turnkey medical and industrial accelerators and sources of synchrotron radiation.

#### 2.2.4.2 The JINR program

One of the express aims of JINR is to promote the development of intellectual and professional capabilities of scientific personnel. To this end its educational program is coordinated and supported as a whole by a special subdivision called the University Center, a formal collaboration of the JINR laboratory, Moscow State University, the Moscow Engineering Physics Institute, and the Moscow Institute of Physics and technology. Elements of training in accelerator science and technology are offered through the Physics Research Facilities Courses[††††].

### 2.2.5. *University programs in Turkey*

As part of an ambitious program of developing accelerator-based science in Turkey, the Turkish Accelerator Center (TAC) has organized the National Summer School Particle Accelerators and Detectors Summer Schools every year since 2005. A primary purpose of the sessions is to introduce students to the fundamentals of accelerator science in preparation for their research in TAC projects. TAC also sends students to regional accelerator schools such as USPAS and the CERN Accelerator School for more advanced training.

### 2.2.6. *Accelerator science in Asia*

Asia possesses a X-ray free electron laser user facility, six third-generation synchrotron light sources, a high intensity proton source, two electron-positron colliders and numerous small accelerator facilities. The need for maintaining a highly trained cadre of accelerator scientists and technologists is evident. This section offers a far from exhaustive view of accelerator training and education in this region.

---

[****] The departmental website is http://www.nsu.ru/exp/ff/kaffu (in Russian).

[††††] http://newuc.jinr.ru/section.asp?id=268

### 2.2.6.1 Accelerator Education in Japan‡‡‡‡

In Japan the High Energy Accelerator Research Organization (KEK) provides education in theoretical and experimental aspects of particle accelerators through its Department of Accelerator Science (Soken-dai). Soken-dai offers a two part course, Introduction to Accelerators, as well as a 3-day summer program for young researchers and college students that provides an experience with real accelerators. The School of High Energy Accelerator Science has three departments – Accelerator Science, Materials Structure Science and Particle and Nuclear Physics – that use the KEK research activities as a basis for graduate education. The Department of Accelerator Science offers research opportunities for both theory and experimental students. Research topics for Ph.D. students include novel acceleration principles and pioneering accelerator technologies, radiation physics, computational accelerator physics, and superconducting magnet and RF technologies. Students carry out research using world-class accelerators and equipment.

### 2.2.6.2 Accelerator education in Australia§§§§

The Australian Collaboration for Accelerator Science (ACAS) supports universities to increase the number and variety of course offerings in accelerator physics.  ACAS also sponsors and dedicated accelerator schools held during non-teaching periods at universities to supplement the material covered in existing courses. The goal is offering a Masters of Accelerator Science as en entry to a career at national accelerator facilities, hospitals or in industry. The main contributors to the teaching of accelerator science are the physics departments at the following universities:

   The University of Melbourne,
   The University of Sydney,
   The Australian National University,
   Monash University, School of Physics,
   University of Wollongong, Centre for Medical Radiation Physics.

### 2.2.6.3 Accelerator training in China

In China the Institute of High Energy Physics (Beijing) supplies graduate courses***** in accelerator science as a department of the Graduate University organized by the Chinese Academy of Sciences. The Department of Engineering Physics of Tsinghua University maintains an active program in accelerator research and education with graduate students at both the Masters and Ph.D. level.

## 3	Regional Schools

At most universities as the number of students at any one time is insufficient to populate courses in specialized accelerator topics (for example, high-power rf-technology, microwave source design, or superconducting magnet technology), a useful approach is for a national or regional laboratory or collaboration thereof, to form an on-going regional school. The archetypes of such regional schools are the U.S. Particle Accelerator School (USPAS) and the CERN Accelerator School (CAS).

### 3.2	The U.S. Particle Accelerator School†††††

USPAS is a national graduate program that provides graduate-level educational courses in the science of particle beams and their associated accelerator technologies that are not generally available to the scientific and engineering communities. The USPAS is governed and funded by an eleven-member consortium of nine national laboratories of the U.S. Department of Energy (DOE) and two NSF university laboratories. The administrative office of USPAS, which performs and coordinates all functions and activities of the School, is funded directly by the DOE's Office of High Energy Physics.  The office is located at

---

‡‡‡‡ The website is http://soken.kek.jp/sokendai_en.
§§§§ See http://accelerators.org.au/

***** See for example, http://english.ihep.cas.cn/et/se/.
††††† The USPAS website is http://uspas.fnal.gov

Fermilab, which provides fiduciary oversight of the School activities for the governing consortium. This governance structure assures that the USPAS programs are highly responsive to the needs of the DOE national laboratories, the DOE and National Science Foundation, U.S. universities and the many students who attend USPAS courses.

The guiding vision of the USPAS is to be an essential partner of U.S. universities and the national laboratories in educating and training the next generation of accelerator scientists and engineers for the challenging accelerators of the future and to advance accelerator science and technology in current research programs. Thus, the USPAS mission is to provide a rigorous and balanced educational program in the science of particle beams and their associated accelerator technologies. A brief history of the School can be found in reference [3].

Since its founding in 1982 by Dr. Melvin Month, the U.S. Particle Accelerator School has continually strived to close the large gap between the need for trained accelerator professionals and the supply provided by the limited university programs in the U.S. Originally organized as symposium style school, the USPAS dramatically changed its format in 1987 to its present, highly successful paradigm in which each session consists of several rigorous courses running in parallel.

Students may register for one full course (≥45 contact hours) or choose two half-courses (≥23 contact hours each) in which each half-course is one week in duration. A full-course offers the equivalent[‡‡‡‡‡] of 3 semester hours of host university credit; each half-course offers the equivalent of 1.5 semester hours of credit. Students may earn graduate academic credit[§§§§§] from the university which hosts the session by completing nightly homework assignments and passing a final exam for their course. All courses run in parallel; hence students can take one full course, or two half-courses. If the hosting university grants half credits, a student may opt to take a single half-course during either week of the program. The percentage of students who take USPAS classes for credit remains high, averaging 63%. In recent years the USPAS has had more than 150 students (of all levels) per session.

The highly varied curriculum of USPAS courses is developed with the advice of a Curriculum Advisory Committee. The Committee is in the process of standardizing the syllabus for all courses that are offered as frequently as every two years. In addition, the host universities require that course descriptions and instructor CVs be submitted roughly one year in advance of the session, to be vetted by their faculty. The USPAS curriculum covers all major aspects of accelerator physics and engineering that are crucial to the user facilities for accelerator-based science.

*Course offerings*

- Fundamentals of accelerator physics, design of storage rings and synchrotrons, linacs, intense beam accelerators, beam optics, spin dynamics.
- Synchrotron radiation sources, free electron lasers, strong field radiation,
- Beam theory, non-linear dynamics, collective effects, beam instabilities,
- Computational methods in beam dynamics, beam optics and electromagnetism,
- Radiation physics and accelerator safety, radiation effects,

*Engineering and technology*

- Experimental techniques of beam physics, microwave measurements and beam instrumentation labs, accelerator vacuum systems, beam manipulation techniques
- RF systems, magnetic systems, superconducting magnets, superconducting RF, superconducting materials, beam sources
- Use of lasers in accelerators, optics-based

---

[‡‡‡‡‡] Some of our hosts are on the quarter system; in that case an equivalent quarter credit is awarded.
[§§§§§] The USPAS does offer one undergraduate course, "Fundamentals of Accelerators."

- diagnostics, optically-based timing systems
- High power electronics, pulsed-power electronics, high power rf-sources, control and feedback systems,
- Radiation shielding and accelerator safety systems,

*Applications and management*

- Accelerator applications in medicine, discovery science, and industry,
- Management of research institutions,
- Project management.

Each year the USPAS offers one or more hands-on laboratory courses in which students learn to use sophisticated instrumentation such as network analyzers and fiber lasers, etc. Full, two-week experimental courses in beam physics at operating accelerators are offered roughly every two years. The most recent of these offerings used the Energy Recovery Linac-based free electron laser at Jefferson Lab. The next such hands-on offering (in the Winter 2013 session) will be at Duke University using the 1 GeV electron storage ring and free electron laser. Unfortunately, due to practical considerations only a dozen students can be handled in such experimental courses on operating accelerators.

Figure 3 displays a histogram of enrollment in courses of various types. Not surprisingly the basic courses are the most popular with students. Typical class enrollment ranges from forty in the undergraduate (Fundamentals of Accelerators) class to several in highly specialized classes. This latter number explains why single universities cannot afford to offer specialty courses, even if appropriate resident or guest faculty members are available to teach.

USPAS faculty members are composed of university faculty (~25%) and of senior researchers from national laboratories (~70%) and industry (~5%) with deep practical experience in specific fields.

As is common at most U.S. universities, at the completion of each course the students provide an evaluation of the course content and of the quality of the instruction. These data provide direct feedback about academic quality to the USPAS Director, Curriculum Advisory Committee, and USPAS Board of Governors as well as to the individual instructors.

The USPAS provides an unparalleled source of continuing education for accelerator physicists, technologists and engineers from its consortium members. Attendees from the national laboratories and partner universities remain a core USPAS constituency. The Fermilab, SLAC and Brookhaven, institutions that historically have had largest accelerator operations (and largest operating budgets), send the largest numbers of participants to USPAS sessions.

*USPAS scholarships*

Thanks to the strong participation and continuing financial backing of the members of the governing consortium, the USPAS is able to offer at least partial scholarship support to nearly all degree-seeking students who register for and are eligible to receive academic credit. Scholarship support is also available to some overseas participants who are not eligible to receive academic credit in the U.S. Scholarship students account for roughly two-thirds of all attendees.

*Some statistics*

The USPAS mechanism allows U.S. universities to offer roughly 800 contact hours annually of rigorous academic instruction in accelerator science and engineering. All together the USPAS has offered over 500 for-credit courses at 46 academic sessions to more than 5000 students. This number includes more than 3400 individuals and more than 1000 students who have taken more than one USPAS class. More than 20% of the USPAS participants from more than 25 countries travel to the U.S. to attend academic sessions. At least 250

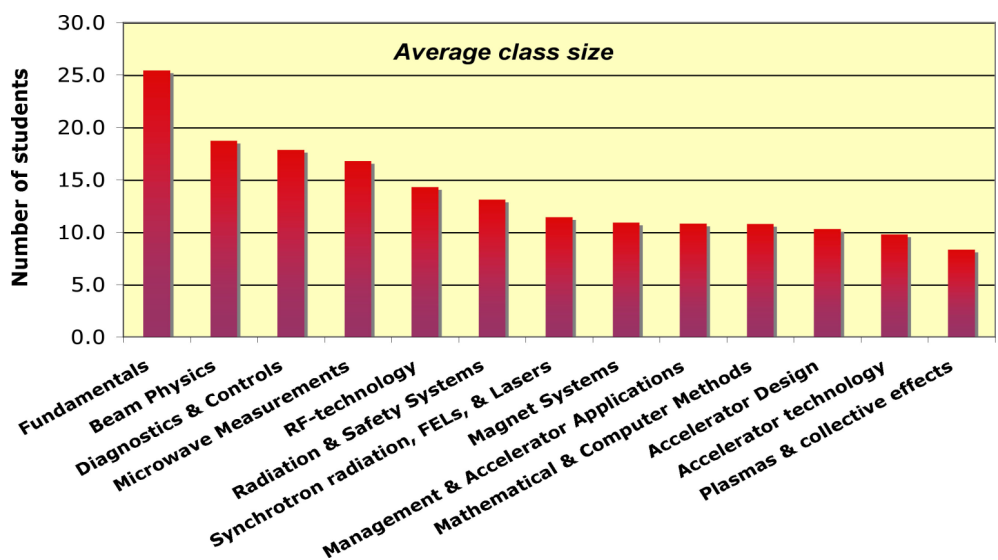

Figure 3. The average class size of USPAS courses by type.

USPAS graduate students have become intellectual leaders in their field, and 25 have returned to USPAS session as instructors. Moreover, 150 USPAS instructors have enrolled in one or more courses at USPAS sessions.

### 3.3 The CERN Accelerator School[******]

In 1983 the European Organization for Nuclear Research (CERN) initiated its own accelerator school, the CERN Accelerator School (CAS), under the leadership of Kjell Johnson and operating with much the same model as the early USPAS symposium-style, topical sessions. The CAS now offers both introductory and specialized training courses for accelerator physicists and engineers twice a year and at periods that do not conflict with the times of the USPAS sessions. The courses consist of lectures and tutorials spread over a period of one to two weeks depending on the topic. Typically each lecturer at a CAS session presents from two to five hours of material. The courses take place in different member states of CERN drawing their participants from the CERN member states as well as other countries.

The CAS pattern of courses is to offer a spring course on a specialized topic with the autumn offering being a general course on accelerator physics. This general course is given at the introductory level in even years and at the intermediate level in odd years. Especially noteworthy is the practice of the CERN School of producing high quality, written proceedings[††††††].

The CAS program of regular courses is augmented every two years by a course on a specialized topic in the framework of the Joint International Accelerator School (JAS) program. The JAS is a collaboration of USPAS, CAS, the Budker Institute for Nuclear Physics in Russia, and KEK in Asia.

In 1985 CAS also initiated a series of annual John Adams lectures. The texts of many of these can be found on the CAS website.

### 3.4 The Joint Universities Accelerator School[‡‡‡‡‡‡]

The Joint Universities Accelerator School (JUAS) was started in 1994 to meet a European demand to train students in the foundations of accelerator

---

[******] The CAS website is http://*cas.web.cern.ch/*.

[††††††] See http://cas.web.cern.ch/cas/Proceedings.html
[‡‡‡‡‡‡] The JUAS website is at http://*juas.in2p3.fr/*.

science, engineering, and technology. The school is organised by the European Scientific Institute[§§§§§§] with the support of fourteen major European universities and the CERN Accelerator School. JUAS offers an intensive programme for students and modular courses for professionals. The JUAS format consists of two five-week courses taught annually by European accelerator specialists. The JUAS curriculum, the constituent course syllabi, and lecturers are recommended and evaluated annually by an advisory board of accelerator scientists and representatives from the supporting European academic institutions.

According to its website, JUAS courses are "intended for students from any recognised European University" and assume no prerequisite knowledge of accelerators. JUAS participants are expected to have "a basic and proven knowledge (4 years at university) of physics or engineering." Students must enrol for a minimum of one five-week course. For staff members and Ph.D. students of laboratories that use particle accelerators and of manufacturing companies that specialize in equipment for accelerators, JUAS does allow part-time participation. As the language of instruction is English and all participants are expected to have a good knowledge of the English language.

The full JUAS program covers a wide variety of topics during ten weeks from January to March. The instructional program includes about 200 hours of lectures, tutorials, guided studies and seminars. The Accelerator Physics covers the following: Introduction to accelerators; Relativity & electromagnetism; Particle optics; Transverse beam dynamics; Longitudinal beam dynamics; Linear accelerators; Space charge; Instabilities; Injection and extraction; Cyclotrons; Synchrotron radiation.

The Technologies & Applications course covers the topics: Introduction to accelerators & components; vacuum systems; radio-frequency engineering; conventional and superconducting magnets; room temperature and superconducting RF systems; beam instrumentation; electron and ion sources; radiation and safety; low energy electron accelerators; production of medical isotopes; radiation therapy; accelerator driven reactor systems; radio-isotopes in medicine; high current proton linacs.

Lectures and tutorials are augmented by site visits and demonstrations plus a practicum at CERN and at the Bergoz Instrumentation Company.

Students at JUAS sessions take examinations under the control of one of the partner universities, which validate the courses. Successful candidates may obtain credits at their home university through the European Credit Transfer System (ECTS)[*******]. JUAS recommends that all students take the examinations, which are mandatory for those students who receive a grant

### 3.5   Joint International Accelerator School[†††††††]

The USPAS and CAS together with the Budker Institute for Nuclear Physics (BINP) in Russian and the KEK Laboratory in Japan have organized a series of accelerator schools focused on specialized topics in accelerator physics and technology. Furthermore, the US-CERN-Japan-Russia Joint International Accelerator School (JAS) aims to strengthen collaborative relationships with the educational programs of the USPAS, CAS, BINP and KEK by working together on an advanced topical course, alternating venues among the four regions.

---

[§§§§§§] According to its web-site "The European Scientific Institute for Applied Physics (ESI) is a non-profit association of 20 European institutions, universities, companies, hospitals and therapy centres from larger Europe. Sited at the "Centre Universitaire de Formation et de Recherche" on the border between France and Switzerland in Archamps, very close to Geneva, ESI was founded in 1994.

[*******] For an official explanation see
http://ec.europa.eu/education/lifelong-learningpolicy/ects_en.htm
[†††††††] Information about past and future Joint International School sessions can be found on the USPAS and CAS websites.

The JAS sessions follow the symposium style of instruction that is used regularly by the CERN School. The First Topical Course in Nonlinear Dynamics was held in Europe 1985 under joint USPAS-CAS sponsorship and was followed the next year by a Course on New Accelerator Methods and Techniques held in Texas. Thereafter, sessions were held every two years through 2002. KEK joined in organizing the Joint School in 1994; BINP joined in 1997.

After a several year hiatus, the organizing institutions have agreed to sponsor the 11th Topical Course in 2011 at the Majorana Center for Scientific Cultural in Erice, Sicily. In 2013, a General Course will be held in Japan to be followed shortly thereafter by a Topical Course on Machine Protection to be held in the U.S.

The JAS sessions foster collaborations among the accelerator communities of the four regions in scholarly work, both writing and teaching. During the school program, participants are encouraged to present and discuss their technical challenges with internationally well-known experts and scientists. In this way the school environment provides a positive atmosphere for mentorship as well as learning. Nine volumes of proceedings have been published.

### 3.6  *Ad hoc topical workshops and schools*

Training in accelerator science may also be offered on an ad hoc basis as required for specific projects or scientific research programs. Typically the goals of such symposium-style schools[‡‡‡‡‡‡‡] are 1) to provide just-in-time training for new project staff and 2) to interest students who may eventually work on or collaborate with the project. This kind of training program is common in many areas of physics.

---

[‡‡‡‡‡‡‡] An early example of such a school was held in 1992 on the site of the then-under-construction Brazilian Light Source in Campinas Brazil. One of the authors (WAB) presented a series of lectures on synchrotron radiation and free electron lasers.

### 3.6.6  *Linear Collider School*

Since 2006 the Global Design Effort of the International Linear Collider (ILC) has sponsored an annual topical school aimed at building a cadre of physicists and engineers for an international TeV-scale lepton collider project, i.e., ILC, the Compact Linear Collider (CLIC), or a Muon Collider. Instructors and students are chosen with balanced representation from the three regions: the Americas, Asia and Europe. Participation is limited to 70 students all of whom receive financial aid (full or partial) for covering the expenses for attending this school, including airfare, lodging, meals, local transportation and school supplies. Applications are limited to graduate students, post doctoral fellows and junior researchers; former attendees are allowed to apply a second time.

The topics covered include the following:

- Overview of TeV-scale future lepton colliders
- Accelerator physics for sources, damping rings, linacs and beam delivery system
- Super conducting and warm RF technology, low-level and high power RF, and beam instrumentation

### 3.6.7  *KoRIA Accelerator School*

In Korea a project to create a facility to study rare isotopes is in its formative stages. This project, KoRIA, will have to assimilate many entry-level researchers who have little familiarity with the constituent technologies. To aid in this process, the project foresees a series topical accelerator schools, the first of which will be held in late 2012.

### 4       Discussion and Conclusions

Throughout the past three decades, education in accelerator science and technology has been carried out in a close, successful partnership among universities, national accelerator laboratories and the regional accelerator schools. Over that same period the accelerator-relevant infrastructure in universities has atrophied considerably – especially in the U.S. Therefore, an important aspect of an accelerator education program in any country should be directed toward strengthening this three-way partnership with the addition of more structured academic programs and more hands-on training opportunities in the major research universities.

Several universities and formalized consortia of universities have expressed recent or renewed interest in developing Masters and Ph.D. degree programs in accelerator physics. However, new or augmented funding is going to have to be available from the national funding agencies to support these programs. Electrical engineering and nuclear engineering departments at some universities also display interests in accelerator education. These departments are important for training students in areas such as control systems, high-power and RF electronics cryogenic engineering or techniques of high-power thermal and radiation load design.

Judging from the steadily rising attendance at USPAS sessions over the past five years – including many applications from students from more than twenty-five countries – student interest in accelerator science and technology has never been higher. Taking advantage of the opportunity these students represent will require an expanded investment in university-based accelerator research and in a new generation of hands-on training instruments. An accompanying expanded program of undergraduate student internships at laboratory research centers would attract some of the most talented undergraduate physics and engineering students into graduate study in accelerator science and technology.

All such educational programs would benefit from expanded outreach to industry, to pre-university school systems, and to the public at large.

### Acknowledgments

One author (WAB) thanks the Board of Governors of the USPAS for its continuing strong support; his work is partially supported by Fermi Research Alliance, LLC under Contract No. De-AC02-07CH11359 with the United States Department of Energy. Author WB thanks the Offices of High Energy Physics, Basic Energy Sciences, and Nuclear Physics of the Department of Energy and the Directorate of Physical and Mathematical Sciences of the National Science Foundation for their strong continuing support of accelerator education in the U.S. Authors SC and AS thank the Science and Technology Facilities Council, UK, for its support of accelerator education and training in the UK. Author SC thanks Prof. Roger Jones of the Cockcroft Institute and University of Manchester and Prof. Carsten Welsch of the Cockcroft Institute and University of Liverpool for their assistance and contributions on the CI Education and Training program and the European Union Initial Training networks respectively. Author AS thanks JAI faculty colleagues in the three universities for the help in describing the JAI training program.